\documentstyle[eqsecnum,amsfonts,aps]{revtex}

\draft

\begin{document}
\title{Asymptotic conditions of motion for radiating charged particles}
\author{James L. Anderson}
\address{Department of Physics, Stevens Institute of Technology, \\
Hoboken, New Jersey 07030}
\maketitle

\begin{abstract}
\newline
\newline
Approximate asymptotic conditions on the motion of compact, electrically
charged particles are derived within the framework of general relativity
using the Einstein, Infeld, Hoffmann (EIH) surface integral method. While
superficially similar to the Abraham-Lorentz and Lorentz-Dirac equations,
these conditions differ from them in several fundamental ways. They are not
equations of motion in the usual sense but rather a set of conditions which
the these motions must obey asymptotically in the future of an initial
starting time. And furthermore, they do not admit the run-away solutions of
these other equations. As in the original EIH work, they are integrability
conditions gotton from integrating the empty-space (i.e., sourceless)
Einstein-Maxwell equations of general relativity over closed two-surfaces
surrounding the sources of the fields appearing in these equations.. No
additional ad hoc assumptions, such as the form of a force law or the
introduction of inertial reaction terms, are required for this purpose. Nor
is there a need for any infinite mass renormalizations such as are required
in other derivations since all integrals are over surfaces and thus finite.
In addition to being asymptotic, the conditions of motion derived here are
also approximate and apply, as do the original EIH equations, only to slowly
moving systems. A `slowness' parameter $\epsilon $ is identified as the
ratio of the light travel time across the system divided by a characteristic
time, e.g., a period. Use is made of both the method of matched asymptotic
expansions and the method of multiple time scales to obtain an asymptotic
expansion in $\epsilon $ and the expansion is carried to sufficiently high
order $\epsilon ^{7}$ to obtain the lowest order radiation reaction terms.
The resulting conditions of motion are shown to not allow run-away motions.
\end{abstract}


\section{INTRODUCTION}

One of the most profound and far reaching consequences of Einstein's general
theory of relativity is the restrictions it places on the motion of the
sources of the fields described by the theory. This is a unique feature of
general relativity and is found in no other field theory. In
electrodynamics, for example, one must postulate not only the form of the
Lorentz force law but also the form of the inertial forces that appear in
the equations of motion. Furthermore, these restrictions, which I have
called conditions of motion, follow from the empty-space field equations of
the theory without the need to introduce source terms into these equations.
As a consequence, many of the well-known difficulties and inconsistencies
associated with point source models are avoided in general relativity. And
finally, these conditions of motion are sufficiently restrictive that they
limit the possible trajectories of an N-body system of compact sources to a
6N-parameter family of curves. In the final analysis, the origin of these
conditions can be traced back to the general covariance of the field
equations of general relativity.\cite{JNG}

Einstein and his co-workers Infeld and Hoffmann (EIH) first derived
approximate conditions of motion for systems of slowly moving ($v/c\ll 1$)
compact gravitating sources up to what is now called the post-Newtonian
approximation in 1939.\cite{EIH} Two subsequent papers on the subject were
published by Einstein and Infeld in 1939 and 1949.\cite{EI} The procedure
used by EIH involved the evaluation of integrals of the field equations over
surfaces surrounding the sources of the fields appearing in these equations.
By using only surface integrals they were thus able to avoid the need to
perform volume integrals over regions containing the sources. Even though
EIH spoke of ``point source'' solutions it is clear that they were really
refering to the form of the solutions in the region exterior to these
sources so that their derivation was applicable to compact sources and black
holes. While the EIH surface integrals were exact their evaluation required
what has come to be called the slow motion approximation. In this
approximation one assumes that the fields and source variables depend on
time scales that are long compared to the light travel time across the
system of sources. In a 1986 paper \cite{JLA1} (referred to hereafter as AI)
I extended this idea by introducing multiple time scales into the theory and
was thereby able to avoid the introduction of the fictitious dipole fields
used by EIH. I also showed how one could make use of the method of matched
asymptotic expansions to include the effects of radiation reaction into the
conditions of motion. These conditions followed as integrability conditions
directly from the empty-space Einstein field equations and were applicable
both to sources with and without strong internal gravity and black holes.
Unfortunately, the EIH papers are referenced only in passing even in
standard texts on general relativity \cite{PGB} and hardly at all in the
literature. And yet these papers contain what I would claim is one of
Einstein's greatest contributions to physics and the only reliable method to
date for deriving conditions of motion.

In this paper I have applied the EIH formalism together with the methods
developed in AI to the case of electrically charged sources up to and
including the order in which radiation reaction terms first appears. Past
attempts to derive an expression for the electrodynamic radiation reaction
force within the framework of Newtonian or special relativistic physics have
been beset with a number of well-known difficulties. The best known of the
equations, the Abraham-Lorentz (AL) equation and its special relativistic
generalization, the Lorentz-Dirac (LD) equation (collectively the ADL
equations) have been the source of continued discussion in the scientific
literature.\cite{TE} - \cite{PAMD}. All derivations of these equations have
required one or more {\it ad hoc} assumptions concerning the form of a force
law, an inertial reaction force or an indeterminate integral. Many
derivations also require some form of infinite mass renormalization and they
all make an arbitrary choice of the retarded over the advanced potential in
constructing the field of a moving charge. Thus Dirac was forced to assume
the form of the indeterminant (in fact, infinite) integrals that appeared in
his derivation.

But perhaps the most serious problem associated with the ADL equations is
the existence of the famous (or infamous, depending on one's point of view)
run-away solutions possessed by these equations. Their existence alone
should have been sufficient evidence that there was something basically
wrong with the whole enterprise. Dirac sought to eliminate these solutions
by imposing the condition that in the asymptotic future all accelerations
should be finite. This condition however leads to the phenomena of
``pre-acceleration'' with its attendant acausality. As a consequence,
several authors have been led to propose alternative equations of motion for
radiating charges.\cite{TCM} Such {\it ad hoc} modifications, motivated by
little more than the desire to avoid the run-away sloution problem can
hardly be considered to be a satisfactory solution to the problem. But more,
it is far from obvious nor has it been shown that any of these equations are
compatible with the restrictions imposed by general relativity on the motion
of sources of the electromagnetic field. And indeed, if one accepts general
relativity as a fundamental theory, one has no alternative than to accept,
as one of its consequences, the restrictions it imposes on the motion of
sources of both the electromagnetic and gravitational fields.

The first derivation of conditions on the motion of electrically charged
sources that did not involve an infinite mass renormalization and/or {\it ad
hoc} assumptions of the type discussed above was that of Infeld and Wallace
(IW).\cite{IW} This derivation applied the EIH procedure to the coupled
Einstein-Maxwell equations and was thus able to avoid the difficulties of
the other derivations. Strangly, this paper is almost never refered to in
the literature. However, in spite of its superiority over other derivations,
the IW derivation did not overcome all of the difficulties of the ALD
equation. IW gave seperate derivations using half-advanced plus
half-retarded and pure retarded potentials and argued that the former was
more ``natural'' even though there was no radiation damping in this case! In
their derivations IW focused on one charge, which they took to be
momentarily at rest, in an external field. After obtaining their conditions
of motion for it from EIH type surface integrals they then performed a
Lorentz boost to obtain conditions for arbitrary velocities. Unfortunately,
dynamical effects involving accelerations can be missed in this way and
indeed, such effects are not present in the IW conditions. They also
assumed, as did EIH, that they had derived equations of motion rather than
asymptotic conditions and hence were unable to rule out run-away solutions.
And finally, theirs was a formal expansion in powers of a parameter $\lambda 
$ which was in the end set equal to unity.

In what follows I will present the details of the derivation of the
conditions imposed on the motion of the sources of electromagnetic and
gravitational fields by applying the EIH method to the coupled
Einstein-Maxwell equations of general relativity. In the next section I will
discuss briefly earlied derivations of the radiation reaction force with
emphasis on the Dirac derivation which is nearest in spirit to the EIH
method so that one can see the essential differences in the two approaches.
Section III will present a derivation of the EIH surface integrals and
Section IV will outline the approximation procedures needed to evaluate
these integrals. Section V contains the details of the derivation and the
paper concludes with a brief summary.

\section{CLASSICAL DERIVATIONS AND THEIR PROBLEMS}

By classical I mean here derivations of the electromagnetic radiation
reaction force that do not use the general theory of relativity. Most of
these derivations attempt to calculate in one way or another a
``self''-force of a charged particle acting on itself. The first derivations
of this kind were given in 1903 by Abraham and 1904 by Lorentz. \cite{ETW}
These early derivations made use of a charge model consisting of a rigid,
spherically symmetric charge distribution $\rho ({\bf x)}$ of radius $a$ and
calculated this force by summing up the interactions of pairs of the
infinitesimal elements comprising the charge. Because of the assumption of
rigidity the results of these calculations were only applicable to
non-relativistic motions. The result of these calculations was a series
expansion of the form

\begin{equation}
{\bf F}_{\text{self }}{\bf =}\sum_{n=0}^\infty {\bf F}_n  \label{2.1}
\end{equation}
where

\begin{equation}
{\bf F}_0=-\frac 43\frac U{c^2}\stackrel{.}{\bf v},
\end{equation}
\begin{equation}
{\bf F}_1=\frac 23\frac{e^2}{c^3}\stackrel{..}{\bf v},
\end{equation}
and 
\begin{equation}
{\bf F}_n\sim \frac{e^2}{n!\,c^{n+2}}\stackrel{(n+1)}{\bf v}\;a^{n-1}.
\end{equation}
Here ${\bf v}$ is the velocity of the charge, $e$ its total charge and $U$
its so-called self-energy which is given by 
\begin{equation}
U=\frac 12\int d^3x\int d^3x^{\prime }\frac{\rho (x)\rho (x^{\prime })}{|%
{\bf x}-{\bf x}^{\prime }|}.
\end{equation}
It is this term which gives most of the trouble in these derivations since
it is proportional to $e^2/c^2a.$ In order to eliminate the higher order
terms in the expression for ${\bf F}_{\text{self}}$ one would like to take
the limit $a\rightarrow \infty .$ But then of course $U$ diverges. There is
however, in my view, an even more serious problem with these derivations. In
order to evaluate ${\bf F}_{\text{self}}$ one must make use of the Lorentz
force law. But that use does not in and of itself justify inserting it into
the right hand side of the Newtonian equations of motion, a procedure I
would call proof by naming. Nor can one be sure that the form of Newton's
inertial force, $m_0\stackrel{.}{\bf v}$ or its special relativistic
generalization, is valid when dealing with charged particles. If one does
proceed in this manner, then one is forced to take for $m_0$, the expression 
\begin{equation}
m_0=m-\frac 43\frac U{c^2},  \label{2.6}
\end{equation}
where $m$ is the physical (finite) mass of the charge, in order to obtain a
finite result in the limit $a\rightarrow \infty $. In this limit the bare
mass $m_0$ is of course negatively infinite.

In order to avoid the non-relativistic restrictions of the AL model, Feynman 
\cite{RPF} suggested using a cutoff procedure in which the Fourier
representation of the retarded Green function is altered by making the
substitution 
\begin{equation}
\frac 1{k^2}\rightarrow \frac 1{k^2}-\frac 1{k^2-\Lambda ^2}
\end{equation}
where $\Lambda $ is the cutoff factor and is ultimately allowed to become
infinite. Using this modified Green function, one can calculate the
self-force of a point charge on itself by evaluating the Lorentz force at
the location of the charge.\cite{SC} The result is finite and is the
relativistic extension of the AL self force with the charge radius $a$
replaced by $1/\Lambda .$. Aside from this feature, this derivation suffers
from the same defects of the AL derivation including the infinite mass
renormalization.

A different approach was taken by Dirac \cite{PAMD} that has some of the
features of the EIH derivation and is reviewed here so that the reader can
compare the two derivations. Dirac considered point charges and defined what
he called the electromagnetic radiation field $F_{\text{rad}}^{\mu \nu }$ to
be 
\begin{equation}
F_{\text{rad}}^{\mu \nu }=F^{+\mu \nu }-F^{-\mu \nu }
\end{equation}
where $F^{+\mu \nu }$ and $F^{-\mu \nu }$ are respectively the retarded and
advanced fields of a such a charge. While these latter fields are infinite
on the charges trajectory, the radiation field is not and is given there by 
\begin{equation}
F_{\text{rad}}^{\mu \nu }=\frac{4e}3\left( \stackrel{...}{z}^\mu \stackrel{.%
}{z}^\nu -\stackrel{...}{z}^\nu \stackrel{.}{z}^\mu \right) .
\end{equation}
where $z^\mu (\tau )$ are the coordinates of the charge expressed as a
function of its proper time $\tau $ and where dots over a quantity indicate
differentiation with respect to $\tau .$ Dirac argued that his definition of
the radiation field was physically reasonable since, at large distances from
the charge and at the correspondingly large times after an acceleration
takes place, the advanced field will be zero. He then assumed the existence
of an energy-stress tensor $_{}$%
\begin{equation}
T^{\mu \nu }=T_{\text{matter}}^{\mu \nu }+T_{\text{em}}^{\mu \nu }
\end{equation}
where $T_{\text{matter}}^{\mu \nu }$ is a matter energy-stress tensor
localized on the particles trajectory and 
\begin{equation}
T_{\text{em}}^{\mu \nu }=\frac 1{4\pi }\left( \frac 14F_{\rho \sigma
}F^{\rho \sigma }\eta ^{\mu \nu }-F^{\rho \mu }F^{\sigma \nu }\eta _{\rho
\sigma }\right)  \label{2.11}
\end{equation}
is the electromagnetic energy-stress tensor. In these equations $\eta _{\mu
\nu }=$ diag(1,-1,-1,-1) is the Minkowski metric tensor and $\eta ^{\mu \nu
} $ is its inverse , Greek indicies take the values 0,...3, and the Einstein
summation convention is used throughout. The total energy-stress tensor is
assumed to be conserved so that 
\begin{equation}
T^{\mu \nu }{}_{,\nu }=0,
\end{equation}
where the comma notation is used to denote differentiation, that is $_{,\mu
}\equiv \partial _{x^\mu }$.

Dirac proceeds by integrating this conservation law over the four-volume
contained within a thin tube surrounding a piece of the charge's trajectory
and converting the result into an integral over the surface of the tube by
means of Gauss's theorem to obtain 
\begin{equation}
\left( \int_{\sigma _1}+\int_{\sigma _2}+\int_S\right) T^{\mu \nu }dS_\nu =0
\label{2.13}
\end{equation}
where $\sigma _1$ and $\sigma _2$ refer to the top and bottom caps
respectively of the tube and $S$ refers to its wall. Since, by assumption,
the support of the matter energy-stress tensor is the charge trajectory it
does not contribute to the integral along the wall. On it Dirac evaluates $%
T_{\text{em}}^{\mu \nu }$ using the total field $F_{\text{tot}}^{\mu \nu }$
given by 
\begin{equation}
F_{\text{tot}}^{\mu \nu }=F_{\text{in}}^{\mu \nu }+F^{+\mu \nu }
\end{equation}
where $F_{\text{in}}^{\mu \nu }$ is the field incident on the charge. The
wall is such that its intersection with each space-like plane whose normal
is parallel to $\stackrel{.}{z}^\mu $ at the point of its intersection with
the charge trajectory is a small sphere of radius $\varepsilon .$ In order
that this construction leads to a well-defined wall it is of course
necessary that $\varepsilon $ be less than the curvature of the trajectory
which is inversely proportional to the acceleration of the charge. Finally
the field on each of these spheres is evaluated in terms of the coordinates
of its instantaneous center. This evaluation cannot be given exactly since $%
F^{+\mu \nu }$ depends on the retarded position of the charge which is given
by an expansion in $\varepsilon $ about its instantaneous position. The
result of these calculations is that 
\begin{equation}
\int_ST_{\text{em}}^{\mu \nu }\,dS_\nu =\int_Q^{Q^{\prime }}\left[ \frac 12%
e^2\varepsilon ^{-1}\stackrel{..}{z}^\mu -e\stackrel{.}{z}^\nu f_\nu {}^\mu +%
{\cal O(\varepsilon )}\right] d\tau  \label{2.15}
\end{equation}
where $Q$ and $Q^{\prime }$ correspond to the points of the trajectory that
intersect the top and bottom caps respectively and where 
\begin{equation}
f^{\mu \nu }=F_{\text{in}}^{\mu \nu }+\frac 12\left( F^{+\mu \nu }-F^{-\mu
\nu }\right) .
\end{equation}
The ${\cal O}(\varepsilon )$ term in (\ref{2.15}) is in fact an infinite
series in $\varepsilon $ similar in form to the $F_n$ in the expression (\ref
{2.1}) above for $F_{\text{self}}$ which Dirac does not evaluate.

In order to complete the derivation it is of course necessary to evaluate
the cap contributions in (\ref{2.13}). But here one encounters two serious
difficulties: one does not know $T_{\text{matter}}^{\mu \nu }$ {\it ab initio%
} and worse, the contribution to the caps from $T_{\text{em}}^{\mu \nu }$ is
divergent. As a consequence Dirac was forced to assume the form of the
contributions from these caps. Since the final answer in any case must be
independent of $\varepsilon $ these contributions must cancel the $%
\varepsilon $ dependent terms in the integral in (\ref{2.15}) and not, as
Dirac asserts, so that the integrals have a definite form in the limit $%
\varepsilon \rightarrow 0$. There is, in fact, no need to take this limit
nor would it make any mathematical sense to do so. These contributions must
also be consistent with the requirement, which follows from (\ref{2.13}),
that the line integral in (\ref{2.15}) depend only on the end points $Q$ and 
$Q^{\prime }$ and hence the integrand in this integral be a perfect
differential. Neglecting the $\varepsilon $ dependent terms which in all
cases must cancel with contributions from the caps this latter condition
tells us that 
\begin{equation}
e\stackrel{.}{z}^\nu f_\nu {}^\mu =\stackrel{.}{B}^\mu 
\end{equation}
where $B^\mu $ is undetermined except for the requirement that $\stackrel{.}{%
z}_\mu \stackrel{.}{B}^\mu =0$ which follows from the antisymmetry of $%
f^{\mu \nu }$ in $\mu $ and $\nu $. The simplest assumption that satisfies
this requirement is that $B^\mu =m\stackrel{.}{z}^\mu $although, as Dirac
pointed out, there are other possibilities. With this assumptionn the
equations of motion become 
\begin{equation}
m\stackrel{..}{z}^\mu =\frac 23e^2\left[ \stackrel{...}{z}^\mu +\stackrel{..%
}{z}^\nu \stackrel{..}{z}_\nu \stackrel{.}{z}^\mu \right] +e\stackrel{.}{z}%
_\nu F_{\text{in}}^{\mu \nu }  \label{2.18}
\end{equation}
which Dirac took to be exact. However, buried within the assumption for $%
B^\mu $ is some kind of infinite mass renormalization to compensate for the
divergent self-energy contribution of $T_{\text{em}}^{\mu \nu }$ to the cap
integration. If, for example, one took, for the matter energy-stress tensor
the Minkowski tensor 
\begin{equation}
T_{\text{matter}}^{\mu \nu }=m_0\int \delta ^4(x-z(\tau ))\stackrel{.}{z}%
^\mu \stackrel{.}{z}^\nu d\tau 
\end{equation}
where $\delta ^4(x)$ is the four-dimensional Dirac $\delta $-function, the
bare mass $m_0$ would contain a negative infinite component similar to the $%
m_0$ in (\ref{2.6}). As the reader will see in what follows, the EIH surface
integral method circumvents this problem by in effect replacing the volume
integrals in the cap contributions by surface integrals which are all
finite. Finally, this equation, like its counterparts obtained from the use
of one or the other radiation reaction force, possesses run-away solutions,
that is, solutions for which the accelerations increases exponentially with
time.

\section{EIH SURFACE INTEGRALS}

The derivation of the conditions of motion makes use of the fact that the
Einstein-Maxwell field equations for the gravitational field $g_{\mu \nu }$
and the electromagnetic field $F^{\mu \nu }\,$contain three dimensional
curls. Therefore, when these equations are integrated over any closed
2-surface these curl terms vanish. When in particular such a surface
encloses a source of these fields the requirement that the remaining surface
terms must be surface independent and vanish leads to conditions on the
motion of this source. To facilitate the construction of these surface
integrals it is useful to write the Einstein-Maxwell field equations in the
form (I use units in which $G=c=1$, Latin indices run from 1 to 3 and Greek
indices from 0 to 3.)

\begin{equation}
U^{\mu \nu \rho }{}_{,\rho }=\Theta ^{\mu \nu }  \label{3.1}
\end{equation}

where

\begin{equation}
U^{\mu \nu \rho }=-U^{\mu \rho \nu }=\frac 1{16\pi }\{(-g)(g^{\mu \nu
}g^{\rho \sigma }-g^{\mu \rho }g^{\nu \sigma })\}_{,\sigma }\text{ ,}
\end{equation}

\begin{equation}
\Theta ^{\mu \nu }=(-g)(T_{\text{em}}^{\mu \nu }+t_{LL}^{\mu \nu }),
\end{equation}
\begin{equation}
\text{(}\sqrt{-g}\,F^{\mu \nu })_{,\nu }=0\,,  \label{3.4}
\end{equation}
and 
\begin{equation}
F[\mu \nu ,\rho ]=0.  \label{3.5}
\end{equation}
Here $g=$det$(g_{\mu \nu }),\,t_{LL}^{\mu \nu }\,$ is the Landau-Lifshitz
pseudotensor $\cite{LL},$ $T_{\text{em}}^{\mu \nu }\,$ is the
electromagnetic energy-stress tensor given by (\ref{2.11}) (in what follows
I will omit the subscript em to simplify the notation), indices are raised
and lowered using $g_{\mu \nu \,}$ in the usual manner and brackets around a
set of indices indicates that the indices are to be antisymmetrized.

Because of the antisymmetry of $F^{\mu\nu}\,$in $\mu$ and $\nu$ and $%
U^{\mu\nu\rho}$ in $\nu$ and $\rho$, it follows that $F^{rs}_{,s}\,$and $%
U^{\mu rs}_{,s}$ are three dimensional curls. As a consequence, integration
of (\ref{3.1}) over a closed spatial 2-surface in a $t=$const. hypersurface
gives the result

\begin{equation}
\oint \text{(}U^{\mu r0}{}_{,0}-\Theta ^{\mu r})n_r\,dS=0  \label{3.7}
\end{equation}
where $n_r$ is an outward pointing unit surface normal. \ Similarly, we get
from equation (2.4)

\begin{equation}
\oint \text{(}\sqrt{-g}\,F^{\text{ }r0}\text{)}_{,0}n_r\,dS=0.  \label{3.8}
\end{equation}

It is these last two equations that are used to obtain conditions of motion
for our charged system. They come from noting that, since these equations
must hold on any two-surface, the sums of the contributions to the integrals
that are surface independent must by themselves be zero. (The surface
dependent terms must in all cases vanish identically.) In practice one
chooses the surfaces to be spheres centered on the sources and sets to zero
those terms that do not vanish identically. (If there are no sources, so
that the fields are divergence free inside the surfaces, then all the
integrals vanish directly as a consequence of the field equations.)

Because one employs in the derivation only surface integrals on which the
fields are everywhere finite, it follows that there are no infinite
integrals that must be renormalized away. If we compare (\ref{3.7}) with (%
\ref{2.13}) we see that the divergent cap terms in the latter equation get
replaced by the finite surface integrals of $U^{\mu r0}{}_{,0}$ and $%
(-g)t_{LL}^{\mu \nu }.$ It is the first of these integrals in fact that
contribute the inertial force terms to the conditions of motion, obviating
the need to postualte them. Furthermore, since the source region is
contained entirely within the 2-surfaces of integration in (\ref{3.7}) and (%
\ref{3.8}), one need only know the fields on these surfaces and not inside.
All that is required is that each surface contain only one source, that
their radii are sufficiently large that the fields on them are weak and
small compared to the distance between sources. Thus one can equally well
deal with compact sources with strong internal gravity, e.g., neutron stars
and even black holes as with sources whose fields are weak everywhere. In
particular, the problems associated with the use of point sources discussed
in Section II do no occur in the EIH procedure.

\section{APPROXIMATION PROCEDURES}

The evaluation of the surface integrals in (\ref{3.7}) and (\ref{3.8})
requires a knowledge of the fields on these surfaces and this can only be
gotten by employing some kind of perturbation scheme. The use of any such
scheme requires that there be one or more small parameters associated with
the system. In the present work one of these is a dimensionless 'slowness'
parameter $\epsilon $ that is of the order of the ratio of the light travel
time across the system of sources to a time scale associated with their
motion, e.g. an orbital period. It is also of the order of the ratio of the
size of the system to the wavelength of the radiation emitted by it and also
the ratio of the velocities of the sources to the velocity of light. Its
precise value is fixed by the parameters that characterize the source system
and its motion. It should be emphasized that this parameter is not an
arbitrary small parameter as EIH took it to be nor can it be taken to be the
inverse of the velocity of light as many authors have done. In addition to
assuming that $\epsilon $ is small compared to unity, I will assume that the
relative strengths of the gravitational and electromagnetic interactions $%
\delta =$mass/charge is small compared to unity and also to $\epsilon $ for
all charges in the system. In this work I have only kept terms of order $%
\delta ^2$ and I have also scaled masses and charges so that $m_A={\cal O}%
(\delta ^2\epsilon ^2)\,$ and $q_A={\cal O}(\delta \epsilon ^2)$. These
restrictions have the effect that the operation of raising and lowering
indices on all tensorial quantities is accomplished with the help of the
flat-space Minkowski tensor.

\subsection{Matched Asymptotic Expansions}

The perturbation methods employed here are the singular perturbation
techniques used in AI to derive approximate conditions of motion for
gravitationally interacting uncharged particles. The first of these is the
method of matched asymptotic expansions which was first used by Laplace in
1805 (see Ref. \cite{AN} for this and other references to this method.) In
general this method seeks to develop approximate solutions to a system of
equations in different regions of the range of the independent variables
using expansions appropriate to each region The two expansions are then
'matched' in an overlap region where both expansions are valid. In order to
obtain approximate solutions to the Einstein-Maxwell equations it is in
principle necessary to use four overlapping regions in all corresponding to
two different matchings, one in time and one in space.

The time matching treats the region in the vicinity of an initial value
surface as a boundary layer called here the fast zone where the solution
depends on the initial value data. The solution in this region is matched to
an asymptotic region in the future of the initial data surface $t=$ const.
called the slow zone. The undetermined parameters and functions appearing in
the solution in this latter region are then determined in terms of the
initial value data by matching to the fast zone. The slow zone can be
thought of as the region in which all initial transients generated by the
initial data have died off.

The other matching is the one used by Burke.\cite{WLB} It matches solutions
in two overlapping spatial regions consisting of an inner or induction zone
and an outer or radiation zone. What is important to us is that, in the
lowest order of approximation the solution in the outer, slow zone is a pure
outgoing solution, that is, a function of a null coordinate $u$ and the
spatial coordinates ${\bf r}$. (In higher orders of approximation this
region will contain both outgoing and incoming radiation due to
backscattering.) It should be emphasized that this form of the solution is
not imposed arbitrarily as it usually is in other treatments. Rather, it is
a consequence of the fact that solutions of the wave equation have the
property that so long as the energy of the field on the initial-value
surface is finite then in the asymptotic future the field is purely outgoing.%
\cite{JLA2}

As we shall see, it is the matching to this outgoing solution that is
responsible for the phenomena of radiation damping and the appearance of
irreversible terms in the conditions of motion to be derived here. Thus we
see that irreversibility in the conditions of motion is a consequence of the
finite energy condition imposed on the initial fields and not, as IW
contended, because of an arbitrary choice in the use of retarded or advanced
potentials. One could choose initial conditions on the fields that would
lead to anti-damping for all times but this would violate the finite energy
condition mentioned above. While the final result of the two approaches
appears to be the same, the one used here is in accord with those used in
other areas of physics where irreversibility arises in the asymptotic future
as a consequence of restrictions on initial conditions.\cite{EAF}

Although in principle one could construct solutions in the fast-time zone to
obtain conditions of motion in this zone, I have not explicitly done so in
this work. In general, the motion of the sources in the fast-time zone will
depend in a very complicated way on the initial values not only of the
source coordinates but those of the fields as well and will almost certainly
not satisfy any unique set of equations of motion. Without this matching the
conditions on the source motion obtained by the other matchings will then
only be valid in the asymptotic future and are therefore not equations of
motion in the usually accepted Newtonian sense. In particular, they cannot
be used to solve initial value problems. I will defer a discussion of the
implications of this state of affairs until after I outline a derivation of
the conditions of motion accurate to radiation damping order.

\subsection{Multiple Time Scales}

In order to construct approximate expansions in the two spatial domains in
the slow-motion domain it is necessary to make use of the method of multiple
time scales. In their original work EIH employed only a rudimentary multiple
time scale formalism. As a consequence they were forced to introduced
fictitious dipole terms at each order of their expansion in $\lambda$ and in
the end require that the sum of all these dipoles vanish. This procedure
appeared somewhat arbitrary and may have cast some doubt on the validity of
their results. In fact, when the multiple time scale formalism is used no
such dipole terms are needed.

As the name implies, the multiple time scale formalism is used when one has
to deal with systems in which phenomena occur on different time scales,
e.g., oscillation and damping. Not only is it useful in separating out these
different types of motion but it is used extensively to avoid the appearance
of secular growth and nonuniformities in various approximation schemes. In
its application one assumes that the dependent variables of a system of
equations depend on the time variable $t$ via a sequence of times, $t_0,t_1,$
$\ldots $ where usually, but not always, $t_0=t$ and $t_n=\epsilon ^nt$, $%
\epsilon $ being a small dimensionless parameter in the problem. In this
case, which is the one that applies in this work, a derivative with respect
to $t$ is replaced by 
\begin{equation}
\partial _t\text{ }\rightarrow \sum_{n=0}\epsilon ^n\partial _{t_n}.
\end{equation}
The dependence on these times is then fixed by the requirement that the
terms responsible for secular growth vanish. (See AI and Ref.\cite{AN} for
simple examples of the application of this method.)

The time scales required in general relativity will depend upon the problem
being considered. If we are dealing with uncharged particles and if $t$ is
the fastest time, characterized by the light crossing time of the system,
then $\epsilon t$ will characterize the Newtonian time scale, $\epsilon
^3t\, $ the post-Newtonian effects such as perihelion advance, $\epsilon ^5t$
the post-post-Newtonian effects etc.. Gravitational radiation is
characterized by $\epsilon ^6t$, sometimes referred to as the 2$\frac 12$%
-post-Newtonian time scale (see Reference 11 for details.) When one is
dealing with charged sources there are at least two different small
parameters that characterize the system, $\epsilon $ and $\delta =m/q$, and
hence the possibility of additional time scales arises. However. I have
assumed that $\delta \ll \epsilon \ll 1$ so as to avoid having to deal with
both electromagnetic and gravitational effects simultaneously. Consequently, 
$t_1=\epsilon t$ will characterize the Coulombian interaction time scale, $%
t_3=\epsilon ^3t$ the Amperian or post-Coulombian time scale and radiation
effects will be characterized by a $t_4=\epsilon ^4t$ time scale in lowest
order.\cite{KST}

\section{\ CONDITIONS OF MOTION}

The construction of approximate solutions of the field equations (\ref{3.1})
and (\ref{3.4}) and their use in evaluating the integrals in (\ref{3.7}) and
(\ref{3.8}) follows the pattern developed in AI. (Here I will use the vector
4-potential $A_\mu $ where $F_{\mu \nu }=A_{\nu ,\mu }-A_{\mu ,\nu }$. As in
AI, I assume the existence of a coordinate map such that ${\frak g}=\sqrt{-g}%
g^{\mu \nu }$ can be expanded up to the order of accuracy needed here in an
asymptotic series of the form 
\begin{equation}
{\frak g}^{\mu \nu }\sim \eta ^{\mu \nu }+\sum_{n\text{,}m}\epsilon
^n\,\delta ^mh_{nm}^{\mu \nu }({\bf r,\,x}_A{\bf ,}q_A,{\bf \,}m_A{\bf )}
\label{5.1}
\end{equation}
where I have assumed that the $h_{nm}^{\mu \nu }$ depend on the multiple
times through their dependence on the source coordinates ${\bf x}%
_A(t_1,\,t_3,\,\ldots ,\epsilon ,\delta )$,\thinspace charges $%
q_A(t_1,\,t_3,\,\ldots ,\,\epsilon ,\,\delta )\,$ and masses $%
m_A(t_1,\,t_3,\,\ldots ,\epsilon ,\,\delta )\,$with $A$ labeling the
sources. One should keep in mind that ${\frak g}^{\mu \nu }$ cannot be
expanded indefinitely in a simple power series since it is not analytic in $%
\epsilon $, reflecting the fact that one encounters $\epsilon ^n\,$ln($%
\epsilon $) terms in higher orders of the approximation.\cite{JLA3} However,
these terms do not appear at the level of accuracy used in this work. In a
like manner I assume that the 4-potential $A^\mu $ can be expanded as 
\begin{equation}
A^\mu \sim \sum_{n,m}\epsilon ^n\delta ^mA_{nm}^\mu ({\bf r},{\bf x}%
_A,\,q_A,\,m_A)  \label{5.2}
\end{equation}
(Recall that, because of the smallness of $\delta $, indices are raised and
lowered here using the Minkowski tensor so that, e.g., $A^\mu =\eta ^{\mu
\nu }A_\nu .)$

Because of the scalings chosen for our sources it follows that the expansion
(\ref{5.1}) starts with ${\cal O}$($\delta ^2$) terms while the expansion (%
\ref{5.2}) starts with ${\cal O}$($\delta $) terms. Since one only needs
terms of this order to obtain an ${\cal O}$($\delta ^2$) overall accuracy in
the conditions of motion I will drop the index $m$ in the above expansions
in what follows. Also, in this order the dependence of the source variables $%
{\bf x}_A,\,q_A,\,m_A$ on $\delta $ can be ignored so that from this point
on I will ignore entirely all $\delta $ dependencies. Then, again because of
the scaling employed, the lowest order in $\epsilon $ fields are $A_2^0$ and 
$h_2^{00}$.

In addition to the field expansions (\ref{5.1}) and (\ref{5.2}), it will
also necessary to make certain assumptions concerning the dependence of the
source variables $q_A$, $m_A$, and ${\bf x}_A$ on $\epsilon $. \ To the
level of accuracy needed here I will assume that the latter two variables
can be expanded as an asymptotic series in powers of $\epsilon $ of the form 
\begin{equation}
m_A\,\sim \epsilon ^2m_{A2}+\epsilon ^4m_{A4}+\cdots
\end{equation}
and

\begin{equation}
{\bf x}_A\sim {\bf x}_{A0}+\epsilon ^2{\bf x}_{A2}+\cdots  \label{5.4}
\end{equation}
where the dots indicate terms that are at least an order of magnitude in $%
\epsilon $ smaller that the terms retained. It should be emphasized that the
powers of $\epsilon $ used in this expansion is not arbitrary but dictated
by approximation itself. As we shall see, the masses $m_A$ will be
automatically expanded as we proceed in the approximation while the charges $%
q_A$ are absolute constants of the motion and that there is no need to
expand them at all.

To obtain equations for the fields $h_n^{\mu \nu }$ and $A^\mu $ from (\ref
{3.1}) and (\ref{3.4}) I will impose the deDonder coordinate conditions 
\begin{equation}
{\frak g}_{\text{,}\nu }^{\mu \nu }=0  \label{5.5}
\end{equation}
and the Lorentz gauge condition 
\begin{equation}
\text{A}_{\text{,}\mu }^\mu =0.  \label{5.6}
\end{equation}

Here too problems arise with the use of these conditions in higher orders of
the approximation and they must be modified to eliminate the appearance ln$%
(r)$ terms in these higher orders.\cite{JLA3}

At this point a difficulty arises. Usually when one substitutes an expansion
of the dependent variables in powers of a small parameter into a system of
equations one simply equates to zero the coefficients of the various powers
of this small parameter. However, in the present case the coefficients $%
h_n^{\mu \nu }$ and $A_n^\mu $ in the expansions (\ref{5.1}) and (\ref{5.2})
themselves depend on $\epsilon $ through the dependence of the source
variables on $\epsilon $. Therefore one cannot simply equate to zero the
coefficients of the powers of $\epsilon $ when these expansions are
substituted into the field equations. This difficulty seems to have been
overlooked in all previous derivations of equations of motion in general
relativity including AI as far as I know. It can however be dealt with in
the following manner: Substitute the expansions (\ref{5.1}) and (\ref{5.2})
into the field equations and set $\epsilon =0 $. It then follows from (\ref
{3.1}) that in the source-free weak-field region$h_2^{00}|_{\epsilon =0}$
satisfies 
\begin{equation}
\nabla ^2h_2^{00}|_{\epsilon =0}=0.  \label{5.7}
\end{equation}

However, since $h_2^{00}$ depends on $\epsilon $ solely through its
dependence on the source variables it follows that it must satisfy (\ref{5.7}%
) for all values of $\epsilon \ll 1$. Since at each stage of the
approximation the unknown field coefficients in the expansions (\ref{5.1})
and (\ref{5.2}) enter the field equations linearly it follows that this
argument holds quite generally and so allows us to proceed to find these
coefficients without regard to the dependence of the source positions on $%
\epsilon $.

The requirement that the sources are characterized only by their charge and
mass is equivalent to requiring that, in the absence of other sources, the
field of each source must be spherically symmetric. \ It follows therefore
that h$_2^{00}$ is given by \ 
\begin{equation}
h_2^{00}=4\sum m_{A2}/r_A.  \label{5.8}
\end{equation}
where the sum is over all sources in the system, i.e., over $A=1$ to $%
N,r_A^i=r^i-x_A^i$ where $x_A^i$ is the $i$th coordinate of the $A$th source
and where, for convenience, we will take the center of mass of our system to
be at the origin of coordinates so that $\sum m_Ax_A^i=0$. (In the case of a
charged black hole, $x_A^i$ is the effective coordinate of its center as
determined by its exterior vacuum solution.) In a like manner the field
equations (\ref{3.4}) allow us to conclude that in the source-free region $%
A_2^0$ satisfies 
\begin{equation}
\nabla ^2A_2^0=0
\end{equation}
and so is given, for spherically symmetric sources, by 
\begin{equation}
A_2^0=\sum q_A/r_A.  \label{5.10}
\end{equation}

It cannot be emphasized too strongly that these solutions hold only in the
exterior of the field sources and are not in any sense `singular'. There are
in fact no singular fields in the usual sense in general relativity as there
are in special relativistic theories for example. The sources can be either
compact matter sources or black holes which are modeled by our assumption
that they have no internal dynamics and are characterized solely by their
mass and charge. The fields (\ref{5.8}) and (\ref{5.10}) are exterior fields
and by Birkhoff's theorem are unique.

It should also be explained why the coordinate $r_A$ appears in the above
solutions and not it's zeroth approximation. Indeed, if we were to
substitute the expansion (\ref{5.4}) into these solutions we could then
expand them about $\epsilon =0$, retain in them only the lowest order terms
and add the remaining higher order terms to the higher order terms in the
expansions (\ref{5.1}) and (\ref{5.2}). In effect, by proceeding in this
manner we have effectively summed partial series that would have resulted
had we not done so. By keeping all of these terms in the lowest order
solutions (and we will continue this practice with all of the higher order
terms) we accomplish two things: Firstly, it greatly simplifies the
calculation due to the effective summing of numerous partial series. But
secondly, and most importantly it enables us to center the surfaces in (\ref
{3.7}) and (\ref{3.8}) on the true centers of the sources rather than on
their approximate centers. Although of course one can use any surfaces one
likes in these equations it again greatly simplifies the resulting
calculations to use the true source centers since doing otherwise results in
spurious multipoles appearing in the higher order approximations.

\subsection{ Coulombian conditions of motion}

With the help of the fields $h_2^{00}$ and $A_2^0$ found above we are now
able to determine the time dependence of the charges and masses appearing in
the fields given above. To do so one needs to evaluate the surface integrals
that appear in (\ref{3.7}) with $\mu =0$ and in (\ref{3.8}) respectively
using these approximate solutions. For convenience one chooses spherical
surfaces centered on the sources with radii large compared to their physical
or Schwarzschild radii in the case of black holes but small compared to the
distance between them. These restrictions on the size of these surfaces are
necessary to insure that the weak field approximation employed here is valid
in their neighborhoods. In carrying out these evaluations one finds two
types of terms - those that depend on the radii of the surfaces and those
that don't. Since the integrals as a whole must be independent of these
radii it follows that the overall coefficients of the various powers of the
radii must vanish identically. It also follows that the radii independent
terms must also vanish since the integral as a whole must vanish. However,
these terms will not vanish identically. The requirement that they do vanish
imposes conditions on the motion of the sources as EIH showed. As a
consequence, one need only pick out those terms in the integrands that are
proportional to the inverse squares of the surface radii and direction
independent and set their coefficients equal to zero to obtain these
conditions.

Substitute then the expansion (\ref{5.1}) into (\ref{3.7}), divide by $%
\epsilon ^2$ and set $\epsilon $ equal to zero. In this way one finds that 
\begin{equation}
\partial _{t_1}m_{A2}=0.
\end{equation}

The surface integral in equation (\ref{3.8}) can be evaluated using $A_2^0$
given by (\ref{5.10}) without the need to set $\epsilon $ equal to zero
since it is linear in $A^0$ and since higher order terms in the expansion (%
\ref{5.2}) will not contribute to the integral because of their angular
dependence. As a consequence we obtain the result that

\begin{equation}
\partial _tq_A=0.
\end{equation}
The $q_A$ are thus absolute constants of the motion. Note that a similar
argument does not pertain in the case of the masses because, in higher
orders of the approximation, $\Theta ^{0r}$ in (\ref{3.7}) will in general
be non-zero.

The next quantity needed to obtain the lowest order conditions of motion is
the lowest order contribution to the field $h^{0r}$. In order to satisfy the
deDonder conditions \ref{5.5} we see that these contributions must be of
order $\epsilon ^3$ since $h_2^{00}{}_{,0}$ is of this order and
differentiation with respect to a spatial coordinate does not change the
order of a quantity. It then follows from the field equations and the
deDonder conditions, using reasoning similar to that which led to the
solution (\ref{5.8}), that $h_3^{0r}$ satisfies 
\begin{equation}
\nabla ^2h_3^{0r}=0
\end{equation}
and therefore is given by 
\begin{equation}
h_3^{0r}=4\sum m_{A2}\,\dot{x}_{Ar}/r_A
\end{equation}
where a dot over a quantity denotes differentiation with respect to $t_1$.

Equation (\ref{3.7}) with $\mu =s $ can now be used in conjunction with the
above fields to obtain the lowest, Coulombian conditions of motion. For this
purpose we need to know $T^{sr}\,$to ${\cal O}$($\epsilon ^4$) and $%
U^{sr0}\, $to ${\cal O}$($\epsilon ^3$). The integral over the spherical
surface surrounding the $A$th source appearing in (\ref{3.7}) will vanish
provided the overall coefficient of the terms in the integrand proportional
to $1/r_A^2$ that remain after dividing by $\epsilon ^4$ and setting $%
\epsilon $ equal to zero vanishes. This will be the case if 
\begin{equation}
m_{A2}\stackrel{..}{x}_{A0}^s=q_AF_{2A}^{s\,0}  \label{5.15}
\end{equation}
where $F_{2A}^{s\,0}\,$is the lowest order electric field at the location of
the $A$th source due to all the other sources and is given by 
\begin{equation}
F_{2A}^{s\,0}=\sum {}^{^{\prime }}q_B\frac{x_{AB\,0}^s}{x_{AB0}^3}\,
\label{5.16}
\end{equation}
where $x_{AB}^s=x_{A}^s-x_{B}^s$ and the prime on the sum indicates that it
is over all $B\neq A$. Note that, at this order of approximation, our
conditions of motion involve only the lowest order contributions to the
masses and positions of the sources. However, unlike in the case of the
fields, we cannot conclude that these equations are valid for non-zero
values of $\epsilon $, i.e., we cannot drop the 0 subscripts in (\ref{5.15})
and (\ref{5.16}). It must be emphasized that the exclusion of $A$ from the
sum in (\ref{5.16}) is not arbitrary but rather is a direct consequence of
evaluating the integrals in (\ref{3.7}). In fact it could not be otherwise
since all integrals are over surfaces and involve only finite quantities
there. It must also be emphasized that the inertial terms in these equations
are not inserted by hand by appealing to Newton's laws of motion but are
already implicitly contained in (\ref{3.7}).

\subsection{Post-Coulombian conditions of motion}

To obtain the next, post-Coulombian, conditions on the motion of the sources
one needs to evaluate the integrals in (\ref{3.7}) with $\mu =s$ to an
accuracy of ${\cal O}$($\epsilon ^6$). The Maxwell stress-energy tensor $%
T^{sr}$ appearing in the integrand can be evaluated to this order from a
knowledge of $A_3^r$ and $A_4^0$. It follows from the field equations (\ref
{3.4}) and the gauge condition (\ref{5.6}) that $A_3^r$ satisfies 
\begin{equation}
\nabla ^2A_3^r=0.  \label{5.17}
\end{equation}
Therefore, as a consequence of the gauge condition, it must have the form 
\begin{equation}
A_3^r=\sum q_A\,\stackrel{.}{x}_{Ar}/r_A.
\end{equation}
These same equations and gauge condition also lead to the equation 
\begin{equation}
\nabla ^2A_4^0-\stackrel{..}{A}_2^0=0.  \label{5.19}
\end{equation}
It follows that $\,A_4^0\,$is given by 
\[
A_4^0=-\frac 12\sum q_A\,\stackrel{..}{r}_A 
\]
$\,$ 
\begin{equation}
=-\frac 12\sum q_A\{\frac 1{r_A}[\,\stackrel{.}{x}_A^rn_A^r]^2+\stackrel{..}{%
x}_A^rn_A^r-\stackrel{.}{x}_A^2\}.
\end{equation}
where $n_A^r=r_A^r/r_A$. No homogeneous solution of equation (\ref{5.17}) is
included here since, by assumption, it would have to be of the form (\ref
{5.10}) and would therefore have the effect of merely redefining the charges 
$q_A$ appearing in these solutions. This expression for $A_4^0$ appears to
contain a term involving the acceleration of the $A$Th source. However the
conditions (\ref{5.15}) can be used to replace the acceleration factors with
Coulomb interaction terms without affecting the overall accuracy of the
expression, i.e., it will still be accurate to ${\cal O}$($\epsilon ^4$).

With these results in hand the evaluation of the second surface integral in (%
\ref{3.7}) over a spherical surface surrounding the $A$Th source to ${\cal O}
$($\epsilon ^6$)$\,$ is straightforward. Again, one only needs to determine
that part of the integral that is independent of $r_A $. One finds in this
way that 
\begin{equation}
\oint T_6^{ms}\,n_{A\,\text{s}}\,\text{dS}_A=\frac 23q_A\,\stackrel{.}{x}%
_{Ar}\,F_{3A}^{mr\,}-q_A\,F_{4A}^{m0}  \label{5.21}
\end{equation}
In this expression $F_{3A}^{r\,m}$ and $F_{4A}^{0m}$ are again the fields at
the location of the $A$th source produced by all of the other sources. \ The
first of these fields is given$\,$by 
\begin{equation}
F_{3A\,rs}=\sum {}^{^{\prime }}(A_{3A\,s,r}-A_{3A\,r,s})=\sum {}^{^{\prime
}}q_B\,\frac 1{x_{AB}^3}(\stackrel{.}{x}_{B\,}^rx_{AB}^s-\stackrel{.}{x}%
_{B\,}^sx_{AB\,}^r).
\end{equation}
Likewise the second field is given by 
\[
F_{4A\,0r}=\sum {}^{^{\prime }}(A_{3A\,r,0}-A_{4A\,0,r}) 
\]

\[
=-\frac 12\sum {}^{^{\prime }}q_B\{\frac 3{x_{AB}^5}x_{A\,B\,}^r(\stackrel{.%
}{x}_{B\,}^sx_{AB\,}^s)^2+ 
\]

\begin{equation}
+\frac 1{x_{AB}^3}(x_{A\,B\,}^rx_{AB\,}^s\,\stackrel{..}{x}%
_{B\,}^s-x_{AB\,}^r\,\stackrel{.}{x}_B^2+\frac 1{x_{AB}}\,\stackrel{..}{x}%
_{B\,}^r\}
\end{equation}
where again the acceleration factors appearing here are to be replaced by
Coulombian interactions with the help of (\ref{5.15}).

It appears that something is wrong with our calculation, since, among other
things we only get 2/3 of the expected Amperian interaction in (\ref{5.21})
and the sign of the second term on the right hand side of this equations
appears to be wrong. We are led to this conclusion however only if we try to
interpret equation (\ref{3.7}) itself as an equation of motion so that we
might be led to expect that $\oint U^{\mu r0}n_r\,dS$ is the 4-momentum of
the source contained within the surface over which the integral is performed
and $\oint \Theta ^{\mu r}n_r$ $dS$ is the 4-force acting on it. In fact
they are not. What happens is that part of the first integral contributes
the missing 1/3 of the Amperian interaction even though the electromagnetic
field does not appear explicitly in its definition. For its evaluation we
need to know $U_5^{mn0}$ which, when use is made of the deDonder conditions (%
\ref{5.7}), is given by 
\begin{equation}
U_5^{mn0}={\frac 1{16\pi }\left(\stackrel{.}{h}_4^{mn}+h_5^{m0}{}_{,n}%
\right) }
\end{equation}
It follows from the field equations (\ref{3.1}) that $h_5^{m0}$ satisfies 
\begin{equation}
\nabla ^2h_5^{m0}-\stackrel{..}{h}_3^{m0}+4F_2^{r0}F_3^{rm}=0  \label{5.25}
\end{equation}
while $h_4^{mn}$ satisfies 
\begin{equation}
\nabla ^2h_4^{mn}-2\delta ^{mn}F_{20r}F_2^{0r}-4F_2^{0m}F_2^{0n}=0.
\label{5.26}
\end{equation}
Thus we see that the electromagnetic field acts, at this order, as a source
of the gravitational field and it is these contributions that in turn supply
the missing 1/3 of the Amperian force.

It is not possible to obtain closed form solutions for either of the above
equations because of the terms that are quadratic in the fields. However,
all that is needed is to determine those parts of the unknown fields $%
h_5^{m0}$ and $h_4^{mn}$ in the neighborhood of a particular source that
will yield surface independent terms in order to evaluate the surface
integrals surrounding this source. The needed terms are found following
procedures developed by EIH and in AI and are given in the appendix. To
obtain the post-Coulombian conditions evaluate the surface integrals in (\ref
{3.7}) with $\mu =s$, take the second derivative with respect to $\epsilon $
and set $\epsilon $ equal to zero. The result of these operations yields the
conditions

\[
m_{A2}\{\stackrel{..}{x}_{A2}^m+2\stackrel{.}{x}_{A0,t_3}^m+\stackrel{.}{x}%
_{A0}^2\stackrel{..}{x}_{A0}^m+\stackrel{.}{x}_{A0}^r\stackrel{.}{x}_{A0}^m%
\stackrel{..}{x}_{A0}^r\}= 
\]
\ \ $\qquad $\ 
\begin{equation}
{q_A\{\partial _{x_{AB0}^r}F_{2A0}^{m0}x_{AB2}^r+F_{4A0}^{m0}+\frac 12%
\stackrel{.}{x}_{A0}^2F_{2A}^{m0}-{F}_{3A0}^{mn}\stackrel{.}{x}_{A0}^n\}}
\label{5.27}
\end{equation}
As in all multiple time approximations, the dependence of $x_{A0}^m$ on $t_3 
$ in this equation is determined by the requirement that the totality of the
terms that would contribute to the secular growth of $x_{A2}^m$ must vanish.

These conditions, together with the conditions (\ref{5.15}), are identical
in form to the first two slow motion approximations one would obtain from
the special relativistic equations of motion for a charged particle of mass $%
m_{A2}$ and charge $q_A$ moving in the field of a collection of other
charges using a multiple time expansion. The second and third terms on the
left side of this equation as well as the second term on the right side are
equivalent to terms in the special relativistic equations that arise from
replacing derivatives with respect to proper time by derivatives with
respect to coordinate time and expanding the factors of $1/\sqrt{1-\text{(}%
\epsilon \stackrel{.}{x}\text{)}^2}\,$ that enter as a consequence of the
replacement. The first term on the right side is equivalent to a term that
is due to the effect of retardation while the last term is the Amperian
force. It should be emphasized again however that the conditions derived
here differ from the special relativistic equations in a fundamental way in
that they are only asymptotic conditions and are not required to be valid at
all times. They are also derived from a more fundamental theory without
having to be postulated {\it ab initio.}

The reader should notice that, without the multiple time formalism, the
first two terms on the left side of equation (\ref{5.27}) would be absent
and we would arrive at two totally different equations, (\ref{5.15}) and (%
\ref{5.27}), for the accelerations $\stackrel{..}{x}_{A0}^m$ and hence a
contradiction. In an attempt to avoid this inconsistency some earlier
treatments expanded the coordinates $x_A$ in a series in $\epsilon $.
However, this procedure by itself will fail without the multiple time
formalism since secular growth, as in the case of periastron advance, will
ultimately lead to a nonuniform expansion.

In arriving at our conditions of motion (\ref{5.27}) we have consistently
ignored surface dependent contributions to our surface integrals since the
sum of these contributions must vanish identically. One such contribution
does deserve special comment however. In evaluating the surface integral (%
\ref{5.21}) one finds a contribution $\frac 23\stackrel{..}{x}_A^m/r_A.$
Aside from a difference in the numerical coefficient, this is the same
contribution appearing in the line intergral in (\ref{2.15}) found by Dirac.
Here it is exactly canceled by a corresponding term coming from first term
in the surface integral in (\ref{3.7}).

\subsection{Radiation reaction}

The effects of radiation reaction can be derived using the matched
asymptotic expansion method of Burke \cite{WLB} and its extension to the
surface integral method in AI coupled with the method of multiple time
scales. Since Burke did not use the surface integral method he had to make
an assumption equivalent to assuming the validity of the geodesic equations
in order to compute the radiation reaction force. Further, the fact that he
did not use multiple time scales meant that his result cannot be considered
to be part of a consistent approximation scheme even though he did obtain
the ``standard'' form of the gravitational reaction force.

To determine the electromagnetic reaction force I will proceed as in AI by
constructing an outer, slow zone solution and match it to an appropriate
inner solution in the neighborhood of a given source. To this end, following
Burke, I introduce outer coordinates $\widehat{x}^\mu =(\epsilon
t,\,\epsilon {\bf r})$. The fields $\widehat{A}^\mu (\widehat{x})=A^\mu (x)$
are again expanded in what is hopefully an asymptotic series in $\epsilon $
where now the coefficients in the expansion are functions of the outer
coordinates. If one imposes the Lorentz gauge condition it then follows from
the field equations that, to order $\delta ^2$, $\widehat{A}^\mu $ satisfies
the flat-space homogeneous wave equation 
\begin{equation}
\text{(}\widehat{\nabla }^2-\partial _{\widehat{t}}^2)\widehat{A}^\mu =0.
\label{5.28}
\end{equation}

Since the inner expansion of the outer solutions of (\ref{5.28}) must match
to the outer expansions of the inner solutions we have already constructed
we need these latter expansions. They are obtained by letting $r\rightarrow
\infty $ while holding $\epsilon $ fixed. Keeping only the lowest order
nontrivial terms needed for this purpose in this limit yields the results 
\begin{equation}
A_2^0\rightarrow \frac Qr+\,P^r\frac{n^r}{r^2}  \label{5.29}
\end{equation}
where

\begin{equation}
P^r=\sum q_A\,x^rA\,,\text{ \ \ \ \ \ }Q=\sum q_A\text{,}
\end{equation}
and

\begin{equation}
A_3^r\rightarrow \frac{\stackrel{.}{P}^r}r.  \label{5.31}
\end{equation}
We see from these outer expansions that the lowest-order contributions to $\,%
\widehat{A}\,^0$ must contain terms with zero- and first-order spherical
harmonics as factors while those to $\widehat{A}\,^r $ must contain
zero-order harmonics.

While the spherical harmonic structure of the outer fields restricts their
form it does not fix it - any linear combination of outgoing and incoming
wave with the appropriate harmonic dependence will do. In most treatments of
radiation it is assumed that only the outgoing wave is present in the wave
zone and indeed such was assumed both by Burke and by me in our respective
previous works. However, as I later showed,\cite{JLA2} if the energy
contained in a wave field on some initial value surface is finite, then in
the asymptotic future of this surface the field will be a pure retarded
wave. (If one is willing to violate this condition then it is possible to
have incoming waves for as long as one wishes into the future.) One could
also have solved the wave equation as a final value problem. If one now
required that the wave energy on a final value surface be finite then the
solution would be a pure advanced wave in the asymptotic past. As a
consequence, we see that the choice here of the retarded wave solution to (%
\ref{5.28}) is not arbitrary. At the same time it means that any results
that depend on this choice such as the about to be derived radiation
reaction force will only by valid in the asymptotic future.

Retarded waves are functions of $\widehat{u}=\widehat{t}-\widehat{r}\,$ and $%
\widehat{r}$. Consequently the ones that have the right angular dependence
are 
\begin{equation}
\widehat{A}\,^0=\frac b{\widehat{r}}+\left\{ \frac{f_r^{\prime }\,(\widehat{u%
},\epsilon )}{\widehat{r}}+\frac{f_r(\widehat{u},\epsilon )}{\widehat{r}%
^{\,2}}\right\} \,n^r  \label{5.32}
\end{equation}
and 
\begin{equation}
\widehat{A}\,^r=\frac{g_{\,r}(\widehat{u},\epsilon )}{\widehat{r}}
\label{5.33}
\end{equation}
where the prime in equation (\ref{5.32}) denotes differentiation with
respect to $\widehat{u}$. The as yet arbitrary functions $f_r(\widehat{u})\,$
and $g_r(\widehat{u})$ appearing in these solutions are not independent.
They must be so chosen that the fields $\widehat{A}\,^0$ and $\widehat{A}%
\,^r $ satisfy the Lorentz gauge condition used in obtaining the wave
equation (\ref{5.28}) from the field equations (\ref{3.4}). The latter will
be the case provided 
\begin{equation}
\text{g}_{\text{ }r}=f_{\,r}^{\prime }
\end{equation}
To determine these arbitrary functions completely we must match the inner
expansions of the outer solutions (\ref{5.32}) and (\ref{5.33}) to the outer
expansions of the inner solutions (\ref{5.29}) and (\ref{5.31}). These outer
expansions are obtained by letting $\widehat{r}\rightarrow 0$ while holding $%
\epsilon $ fixed and are given by 
\begin{equation}
A\,^0\rightarrow \frac b{\epsilon r}+\left\{ \frac{f_r(\epsilon \,t)}{%
(\epsilon \,r)^2}-\frac 12\stackrel{..}{f}_r(\epsilon \,t)+\frac 13\stackrel{%
...}{f}_r(\epsilon \,t)\epsilon \,r+\cdots \right\} n^r  \label{5.35}
\end{equation}
and 
\begin{equation}
A\,^r\rightarrow \frac{\text{g}_{\,r}(\epsilon \,t)}{\epsilon \,r}-\dot{g}%
_{\,r}(\epsilon \,t)+\cdots .  \label{5.36}
\end{equation}
where again a dot over a quantity denotes differentiation with respect to $%
t_1=\epsilon \,t$. The inner and outer expansion will match if we take 
\begin{equation}
b=\epsilon ^3Q\text{ \ \ \ \ and \ \ \ }f_{\text{ }r}(\epsilon \,t)=\epsilon
^4P^r(\epsilon \,t).
\end{equation}

In the inner expansion (\ref{5.35}) the first two terms match to the first
two terms in the outer expansion of $A_2^0$, the third term matches to the
first term in the outer expansion of $A_4^0$ and the first term of the inner
expansion (\ref{5.36}) matches to the first term in the outer expansion of $%
A_3^r$. However, the time-odd terms in these expansions, the fourth term in (%
\ref{5.35}) and the second term in (\ref{5.36}), do not match to terms in
either of these outer expansions. Since both of these terms are solutions of
the homogeneous Laplace equation a matching can be achieved by adding
solutions of the Laplace equation to our inner solutions whose outer
expansions match to these time-odd terms. The required terms are 
\begin{equation}
A_5^0=\frac 13\sum q_A\,r_{A\,}^r\,\stackrel{...}{x}_A^r\,
\end{equation}
and 
\begin{equation}
A_4^r=\sum q_A\,\stackrel{..}{x}_{A\text{ }}^r
\end{equation}

The above potentials can now be used to evaluate additional surface
independent contributions to (\ref{3.7}) with $\mu =s$. These lowest order
contributions to the second integral in this equation are ${\cal O}$($%
\epsilon ^7$) and, for a sphere surrounding the $A$th source, are given by 
\begin{equation}
\oint_{\,\,\,\,A}T_7^{sr}n_{A\,}^r\,dS_A=\frac 23q_A\sum q_B\,\stackrel{...}{%
x}_{B\,.}^s.\text{ }
\end{equation}
Note in particular that in this expression the sum is over all of the
sources including the $A$th source itself and thus includes, for the first
time in these calculations, the effect of a self action. We shall see that
this self action is similar to the radiation reaction in the ALD equations.

There is also an ${\cal O}$($\epsilon ^7$) surface independent contribution
to the first integral in equation (\ref{3.7}) and is given simply by 
\begin{equation}
\left[ \oint_{\text{ \ \ \ A}}U_{,0}^{sr0}n_{A\,}^r\,dS_A\right]
_7=2m_{A0}\,\partial _{t_4}\stackrel{.}{x}_{A\,}^{\,s}\,.
\end{equation}
where $t_4=\epsilon ^4t$. The ${\cal O}$($\epsilon ^7$) conditions of motion
are then gotten by dividing equation (\ref{3.7}) by $\epsilon ^2$,
differentiating five times with respect to $\epsilon $ and finally setting $%
\epsilon $ equal to zero. \ The result of these operations is 
\begin{equation}
2m_{A0}\,\partial _{t_4}x_{A\,0}^{\,s}=\frac 23q_A\sum q_B\,\stackrel{...}{x}%
_{B\,0.}^{\,s}  \label{5.42}
\end{equation}

The right side of (\ref{5.42}) contains the first time irreversible term in
our conditions of motion. Its form and sign is a direct consequence of
having solved our wave equations as an initial value problems, leading to
the retarded solution in the asymptotic future. If they had been solved as a
final value problem so that the solution would have been an advanced wave in
the asymptotic past then one would have obtained a result similar to (\ref
{5.42}) but with the sign of the radiation reaction term reversed, thus
maintaining the complete symmetry of the basic field equations (\ref{3.1})
and (\ref{3.4}). Put another way, the arrow of time points in both
directions. Of course, if one were to replace $t$ by $-t$ in the past the
advanced solution would transform into a retarded solution and the sign of
the reaction term would be reversed.

\section{ALD and EIH COMPARED}

We have now arrived, starting from the empty-space Einstein-Maxwell
equations (\ref{3.1}), (\ref{3.4}) and (\ref{3.5}) and making only the
assumption of spherical symmetry for the sources of the gravitational and
electromagnetic fields, at a set of constraints (\ref{5.15}), (\ref{5.27})
and (\ref{5.42}) on the motion of these sources. They are applicable only to
slowly moving sources and then only in the asymptotic future of some initial
value surface. As such, they stand in stark contrast to the ALD equations of
motion (\ref{2.18}) which are held to be exact and to hold for all times. If
one were to forget the problems associated with their derivation one would
surely favor them over the EIH equations even on esthetic grounds. Except
for one problem. The ALD equations admit physically unacceptable solutions -
the so-called run-away motions. The EIH conditions on the other hand do not
suffer from this defect as we shall see.

The source of the run-away solutions to the ALD equations is of course the
third derivative terms in the expression for the radiation reaction force
appearing in (\ref{2.18}). Third derivatives also appear in the radiation
reaction term in (\ref{5.42}). However there is an essential difference in
the two cases. In (\ref{5.42}) the third derivative is with respect to $t_1$
and not to the single time variable $t$ appearing in (\ref{2.18}). The
dependence of the source coordinates on $t_1$ (and $t_3$) is fixed by the
Coulombic and post-Coulombic conditions (\ref{5.15}) and (\ref{5.27}).
Equation (\ref{5.42}) serves to determine the dependence of the source
coordinates on damping time $t_4$ and contains only first derivatives with
respect to this variable. As a consequence, the motions allowed by the EIH
conditions are all well behaved. That there are no run-away motions allowed
in general relativity follows from the work of Gibbon, Hawking, Horowitz and
Perry \cite{GG} who have shown that the Bondi mass of a system of charged
masses and/or charged black holes is positive definite and the fact that if
the Bondi mass is initially finite, its time derivative is negative
semi-definite. The Bondi mass thus appears to play the role of a Liapunov
function and its existence therefore strongly suggests that these systems
are asymptotically stable.

To better understand the fundamental difference between the ALD and EIH
equations let us apply them to a simple problem - a non-relativistic
oscillator. In this case the ALD equation reduces, for an electron, to 
\begin{equation}
\stackrel{..}{x}+\omega _0^2x=\tau \stackrel{...}{x}  \label{6.1}
\end{equation}
where $\tau =\frac 23e^2/mc^3$ is the light travel time across the classical
radius of the electron. In keeping with the assumption of non-relativistic
motion I assume that $\omega _0\tau \equiv \epsilon \ll 1$. Finally,
introduce a new independent variable $t^{\prime }=\omega _0t$ to obtain the
equation 
\begin{equation}
\stackrel{..}{x}+x=\epsilon \stackrel{...}{x}.
\end{equation}
The solution will be proportional to an exponential $e^{\lambda t}$ where $%
\lambda $ satisfies 
\begin{equation}
\lambda ^2+1-\epsilon \lambda ^3=0.  \label{6.3}
\end{equation}
Rather than write out the exact solution for $\lambda $ one can construct a
power series solution in $\epsilon .$ One finds that 
\begin{equation}
\lambda _{\pm }=\pm i-\frac 12\epsilon +{\cal O}(\epsilon ^2)  \label{6.4}
\end{equation}
There is a third solution given by 
\begin{equation}
\lambda =\frac 1\epsilon +{\cal O}(\epsilon )  \label{6.5}
\end{equation}
which is the run-away mode. It has the peculiar feature that $\lambda
\rightarrow \infty $ as $\tau \rightarrow 0$, clearly indicating that
something is wrong with our equation. Dirac sought to remedy the situation
by requiring that the initial acceleration, which must be specified along
with the initial position and velocity since we are dealing with a third
order equation, should be such as to rule out the run-away solutions. \cite
{CJE} Aside from its {\it ad hoc} character, this requirement was shown to
lead to acausal behavior. In the present case it would amount to taking the
amplitude of the run-away mode to be zero.

Consider now the solution of the same problem using the EIH conditions of
motion. In this case (\ref{5.15}) would have the form 
\begin{equation}
m_2\stackrel{..}{x}+\omega _0^{\prime }{}^2x=0
\end{equation}
where dots now indicate differentiation with respect to $t_1$ and a prime is
used to differentiate EIH quantities from their corresponding ALD
counterparts. Motions allowed by these conditions are of the form 
\begin{equation}
x=A_{\pm }(t_4)\text{exp}(\pm i\omega _0^{\prime }t_1)
\end{equation}
where the amplitudes $A_{\pm }$ are now taken to be functions of $t_4$. When
this result is substituted into (\ref{5.42}) we obtain the equation 
\begin{equation}
2i\omega _0^{\prime }{}^2\partial _{t_4}A_{\pm }=-\tau ^{\prime }\omega
_0^{\prime }{}^3A_{\pm }
\end{equation}
where $\tau ^{\prime }$ is the same function of the scaled charge and mass
of the electron as $\tau $ is of their unscaled counterparts. With the
scalings introduced at the beginning of Section IV we have $\epsilon
^{\prime }{}^2\tau ^{\prime }=\tau $. Taking $\epsilon ^{\prime }\omega
_0^{\prime }=\omega _0$ we find that 
\begin{equation}
A_{\pm }=A_{0\pm }\text{exp}(i\tau \omega _0^{\prime }{}^2t).
\end{equation}
Thus we obtain two damped modes corresponding to the two roots (\ref{6.4})
of (\ref{6.3}). But we do not obtain a motion corresponding to the run-away
mode (\ref{6.5})

\section{SUMMARY}

The conditions of motion derived above together constitute asymptotic
conditions on the motion of charged, spherically symmetric, compact, slowly
moving sources accurate to ${\cal O}$($\delta ^2$,$\epsilon ^7$). \ Given
the current positions and velocities of a collection of such sources
sufficiently far into the future of an initial startup time they will
predict their motion for at least a finite time into the future to this
order of accuracy. \ They were derived from the source-free Einstein-Maxwell
field equations without additional assumptions concerning force laws or the
form of inertial forces. \ Their derivation did not require assumptions
concerning which parts of the fields act on which sources as does the
derivation given by Landau and Lifshitz,\cite{LL} for example, nor did it
require the renormalization of divergent integrals such as in the Dirac
derivation. \ And these conditions do not admit non-physical solutions which
violate the approximations used in their derivation. In this connection,
Landau and Lifshitz are, to the best of my knowledge, the only authors who
suggest that the existence of run-away solutions is evidence for the limited
applicability of the radiation reaction force. \ But then, like other
authors, they inexplicably ascribe this state of affairs to the
renormalization of the infinite electromagnetic self energy of point sources
when in fact no such renormalization is necessary.

We see thus that the outstanding problems of classical electrodynamics, the
need for infinite renormalizations and the run-away solutions, find their
resolution within the framework of the general theory of relativity. \ In
addition we see how irreversible effects, here radiation damping, arise from
reversible fundamental equations. \ Rather than having to assume an outgoing
wave solution in the slow zone valid for all times it is sufficient to
assume only that the initial conditions are physically reasonable, that is,
that the initial wave energy is finite.

One can ask: is it possible to do better than what has been accomplished
here? One can of course go to higher order in $\epsilon $ in the ${\cal O}%
(\delta ^2)$ approximation, obtaining thereby higher order special
relativistic corrections. One can also go to higher order in $\delta $ to
obtain corrections describing electro-gravitational effects.But can one
derive equations of motion for these sources rather than conditions of
motion? And can one at least derive conditions of motion for arbitrarily
moving sources that are special relativistic invariant. The answer to all of
these questions is, I believe, most likely no. In the fast zone the motion
of the sources will depend in a complicated way not only on their initial
positions and velocities but on the initial values of the fields as well.
Since during this period there will not in general be any small parameters
associated with the motion one would have to solve the field equations
exactly in order to evaluate the EIH surface integrals and that, so far, has
not been done. It is only if one waits long enough for all of the initial
transients in the motion to die down that one can expect to derive
relatively simple conditions of motions. \ For a similar reason one cannot
derive special relativistic equations or conditions of motion. \ All
attempts to do so in the past have been singularly unsuccessful, both
literally and figuratively. \ One of course should not feel to badly about
this state of affairs since there really does not exist a self-consistent
special relativistic description of interacting charged bodies.

To conclude, I want to argue that the derivations of EIH and those here shed
an entirely new light on the meaning of equations of motion which lie at the
very heart of \ classical physics. \ It is not just the fact that conditions
of motion can be derived from the empty-space field equations of general
relativity that is important but even more so is their very nature. In
contrast to equations of motion they are both approximate and asymptotic,
they do not hold for all times and they cannot be used to solve initial
value problems. \ This of course does not mean that everything done in the
past in classical mechanics is wrong because it used these conditions of
motion as equations of motion since their application has always been
compatible with the approximations made here. But it does mean, I would also
argue, that, in a very real sense, all of classical mechanics and
electrodynamics can be looked on as a verification of general relativity
since one does not have to postulate the conditions of motion which underlie
these theories but can derive them from general relativity.

\section*{ACKNOWLEDGMENT}

I would like to thank Joshua N. Goldberg for several very helpful
discussions and suggestions for improving the presentation.

\appendix
\section*{}
In this appendix I will outline the method of determining the field
components needed to derive the post-Coulombian conditions of motion (\ref
{5.27}). The first component needed is $h_4^{00}$, which satisfies 
\begin{equation}
\nabla ^2h_4^{00}=\stackrel{..}{h}_2^{00}-2F_{20r}F_{20r}.
\end{equation}
This equation can be solved exactly by 
\[
h_4^{00}=-(A_2^0)^2 
\]
\begin{equation}
-2\sum m_{A2}[\frac 1{r_A}Y_{A\,}^{rs}\stackrel{.}{x}_{A\,}^{\,r}\,\stackrel{%
.}{x}_{A\,}^{\,s}+Y_{A\,}^r\stackrel{..}{x}_{A\,}^{\,r}]+4\sum \frac{m_{A2}}{%
r_A}  \label{A2}
\end{equation}
where 
\begin{equation}
Y_{A\,}^{rs}=n_{A\,}^rn_{A\,}^s-\frac 13\delta ^{rs}\,\text{ \ and \ }%
Y_{A\,}^r=n_{A\,}^r,\text{ \ \ \ }n_{A\,}^r=\frac{r_{A\,}^r}{r_A}
\end{equation}
are spherical harmonics of orders two and one, respectively. The coefficient 
$m_{A4}$ in the homogeneous term in the solution (\ref{A2}) is fixed by (\ref
{3.7}) with $\mu =0$ using the ${\cal O(\epsilon }^5)$ terms. The surface
independent terms can be made to vanish in this order if we take $\partial
_{t_3}m_{A0}=0$, set 
\begin{equation}
m_{A4}=\frac 12m_{A2}\stackrel{.}{x}_{A0}^2+\frac 12\sum {}^{^{\prime }}%
\frac{q_{B\,0}}{x_{AB\,0}}
\end{equation}
and include the constant of integration in a redefinition of $m_{A2}$.

The next quantity we need to determine is $h_4^{mn}$. Unfortunately, there
does not exist a closed form solution of equation (\ref{5.26}) for these
field components. However, all we need to know are those parts of $h_4^{mn}$
required for the evaluation of the surface independent terms of the first
integral in equation (\ref{3.7}) with $\mu =s\,$ over a sphere surrounding
the $A$th source. These parts can be obtained by expanding the right hand
side of equation (\ref{5.26}) in a Laurent series in $r_{A}$ and keeping
those terms that generate these desired parts. One finds in this way that 
\begin{equation}
h_{4A}^{rs}\simeq 2\,m_{A2}\{\delta ^{rs}n_{A\,}^i\,\stackrel{..}{x}%
_{A\,0}^i-n_{A\,}^r\,\stackrel{..}{x}_{A\,0}^s-n_{A\,}^s\,\stackrel{..}{x}%
_{A\,0}^r\}+\frac{b_A^{rs}}{r_A}
\end{equation}
where the $\simeq $ indicates equality modulo terms that do not contribute
surface independent terms. The coefficient $b_A^{mn}$ in the homogeneous
term is fixed by the imposition of the deDonder conditions (\ref{5.7}) and
is given by 
\begin{equation}
b_A^{rs}=4m_{A2\,}\stackrel{.}{x}_{A\,0}^{\,r}\,\stackrel{.}{x}_{A\,0}^{\,s}
\end{equation}

In a similar manner, the needed parts of $h_{5A}^{m0}$ are obtained from
equation (\ref{5.25}) and are found to be given by

\[
h_{5A}^{0n}\simeq \text{ }2q_An_{A\,}^r\,F_{3A}^{r\,n}+2q_A(\stackrel{.}{x}%
_{A0}^{\,n}n_A^r-\stackrel{.}{x}_{A0}^{\,r}n_A^n)\,F_{2A}^{r\text{ }0}- 
\]
\[
-2m_{A0}\frac 1{r_A}(n_A^rn_A^s-\frac 13\delta ^{rs})\stackrel{.}{x}%
_{A0}^{\,r}\stackrel{.}{x}_{A0}^{\,s}\stackrel{.}{x}_A^{\,n}-2q_An_A^r\,(2%
\stackrel{.}{x_{A0}^r}\,F_{2A}^{n0}+\stackrel{.}{x}_{A0}^{\,n}%
\,F_{2A}^{r0})+ 
\]
\begin{equation}
+\frac{f_A^n}{r_A}
\end{equation}
The last term is a homogeneous term which must again be determined by the
imposition of the deDonder coordinate conditions and which in turn requires
the knowledge of the $h_4^{00}$ found above. When the deDonder conditions
are applied one finds that 
\begin{equation}
f_A^n=\frac{10}3m_{A0}\stackrel{.}{x}_{A0}^2\stackrel{.}{x}_{A0}^{\,n}.
\end{equation}

When these expressions for $h_{4A}^{rs}$ and $h_{5A}^{m0}$ are used to
evaluate $U_{5A}^{mn0}$ and the surface independent terms of the first
integral in equation (\ref{3.7}) together with equation (\ref{5.21}) one
obtains the post-Coulombian conditions (\ref{5.27})$\,$as described in the
text.

\end{document}